\documentclass{eas}
\usepackage{graphicx}
\usepackage{multicol}
\usepackage{natbib}
\begin{document}
\title{On the shape of core overshooting in stellar model computations, 
       and asteroseismic tests} 
\runningtitle{The overshooting properties of KIC\,10526294}
\author{Ehsan Moravveji}\address{Institute of Astronomy, 
Celestijnenlaan 200D, B-3001 Leuven, Belgium}
\thanks{E-mail: Ehsan.Moravveji@ster.kuleuven.be}
\begin{abstract}
Slowly pulsating B stars (SPB) and $\gamma$ Dor stars pulsate in high-order 
gravity (g-) modes.
The frequencies of g-modes are sensitive to the detailed structure and evolution 
history of stars having convective cores.
Receding convective cores in OB-type stars leave behind a chemically inhomogenous 
$\nabla_\mu>0$ radiative zone.
Once a g-mode has radial nodes near the boundaries of these layers, the mode 
gets trapped and its period deviates from asymptotic period spacing. 
Careful study of such trapped modes allows constraining the extent
of such layers by fitting individual pulsation frequencies.
We employ 19 consecuitve dipole g-modes of a very rich \textit{Kepler} SPB pulsator, 
KIC\,10526294, to demonstrate the power of mode trapping in B-stars in studying
the thermal and chemical stratification in the overshooting layer.
\end{abstract}
\maketitle
\section{Introduction}\label{s-intro}
Once OB-type stars reach the zero-age main sequence (ZAMS), they immediately develop 
fully mixed convective cores, which already attains the largest possible mass and 
extent at ZAMS.
During the evolution, the drop in center hydrogen ($X_{\rm c}$) is 
accompanied by the decrease in radiative temperature gradient $\nabla_{\rm rad}$.
Consequently, the convective core shrinks monotonically.
This leaves behind CNO enriched material and a steep increase in hydrogen abundance 
outside the boundary.
Therefore, $\nabla_\mu=(\partial\ln\mu/\partial\ln p)>0$;
here, $\mu$ is the mean molecular weight, and $p$ is the total pressure.
The boundary layer mixing induced by the overshoot of convective plumes into 
the stable radiative layers can partially mix the hydrogen, helium and metals
outside the core, and reshape the $\nabla_\mu$ profile there.

The problem of the boundary layer mixing is still not understood well.
Recently, \citet{viallet-2015-01} proposed that the overshooting mixing depends on the 
local P\'{e}clet number -- defined as the local ratio of the thermal to viscous timescales -- 
in the overshooting layer:
Closer to the core boundary, the convective plumes are predicted to be quasi-adiabatic, 
increasing the size of the fully mixed core.
Furhter away, the P\'{e}clet number drops steeply, and the overshooting layer becomes
photon-dominated, and the mixing proceeds diffusively.
These two influence the shape of $\nabla_\mu$ just ouside the core,
and it propagates directly into the pulsation equations through the definition of
the Brunt-V\"{a}is\"{a}l\"{a} frequency $N^2_{\rm BV}$
\begin{equation}\label{e-brunt}
N^2_{\rm BV} = 
 \frac{g\delta}{H_p} \left(\nabla_{\rm ad}-\nabla+\frac{\phi}{\delta}\nabla_\mu\right),
\end{equation}
where $H_p$ is the local pressure scale height, 
$\nabla=(\partial\ln T/\partial\ln p)$ is the actual temperature gradient
$\nabla_{\rm ad}$ is the adiabatic temperature gradient,
$\delta=(\partial\ln\rho/\partial\ln T)_{P,\mu}$ 
and 
$\phi=(\partial\ln\rho/\partial\ln \mu)_{P,T}$. 
The local thermal and compositional stratifications are encapsulated in $\nabla$ and
$\nabla_\mu$ in Eq.\,(\ref{e-brunt}).

The Brunt-V\"{a}is\"{a}l\"{a} frequency is explicitly present in the adiabatic and
non-adiabatic linearized oscillation equations \citep{unno-1989-book,aerts-2010-book}
Thus, even a slight modification to the $\nabla_\mu$ by mixing (through
overshooting or extra diffusive mixing) in the radiative part of the star influences
the eigenfunctions and eigenfrequencies, and exhibits measurable fingerprints in 
period spacing $\Delta P=P_{n+1}-P_n$, where $P_n$ is the period of a dipole 
g-mode of radial order $n$.
Asteroseismology of heat-driven g-mode pulsators opens a direct window to study 
the physical properties of the fully-mixed core, and the partially homogeneous
layer on top of the core \citep{miglio-2008-01}.

Recently, \citet{papics-2014-01} studied the SPB star 
KIC\,10526294, and identified 19 dipole g-modes from the triplet structure 
around each mode.
\citet[][MAP15]{moravveji-2015-01} carried out a detailed forward seismic modelling of the 
dipole zonal frequencies, and \citet{triana-2015-01} inferred the internal 
rotation profile by inverting its rotational splittings.
We demonstrate how trapping of high radial order g-modes in the $\nabla_{\rm\mu}$-layer
is used to confine the size of this layer.
We also argue that our current treatment of the overshoot mixing requires a major revision.

\section{Simplified picture of mixing in B stars}\label{s-profiles}

\begin{figure}
\begin{minipage}{0.49\textwidth}
\includegraphics[width=\columnwidth]{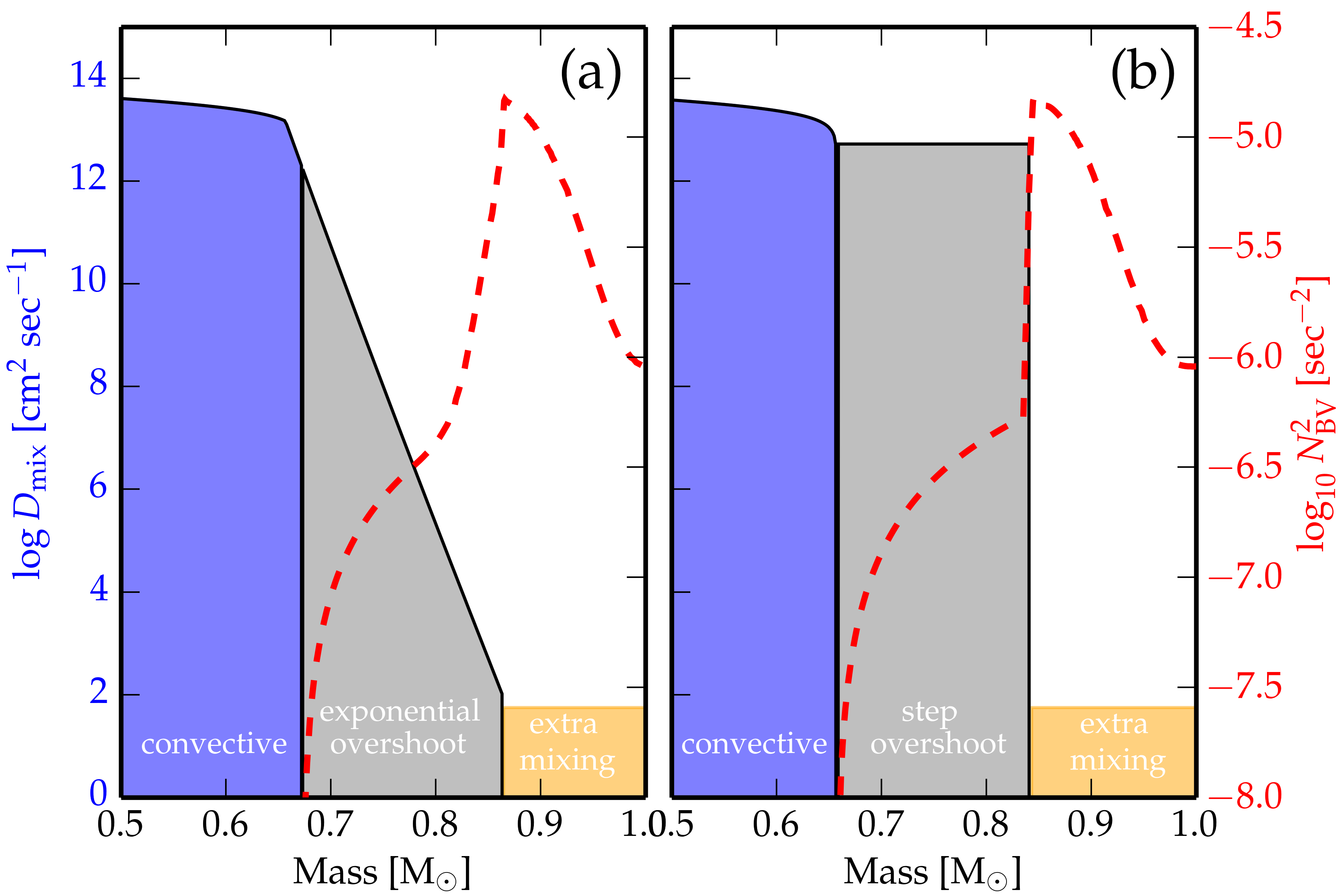}
\end{minipage}
\hspace{0.01\textwidth}
\begin{minipage}{0.49\textwidth}
\includegraphics[width=\columnwidth]{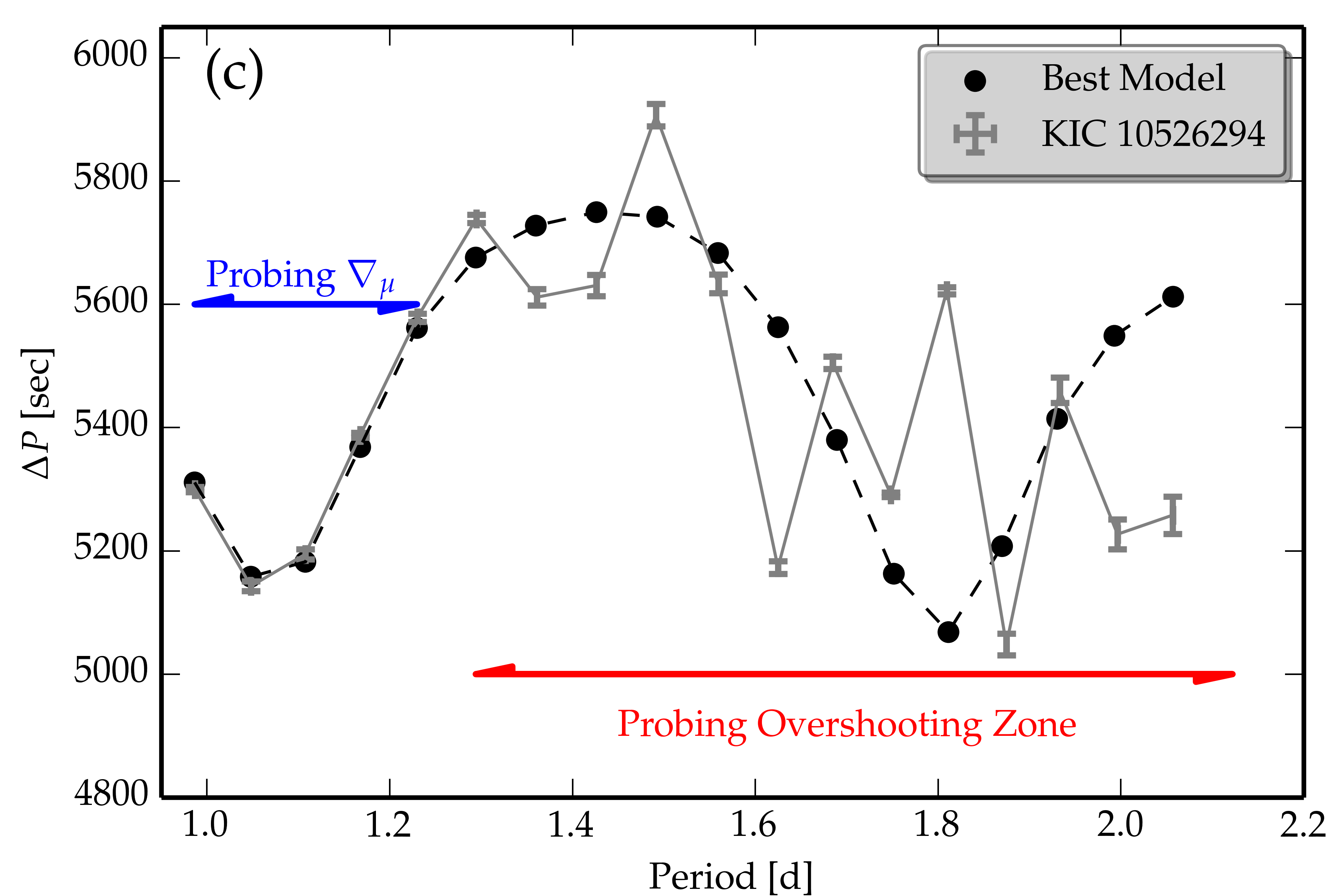}
\end{minipage}
\caption{The convective (blue), overshooting (grey) and extra diffusive mixing (orange)
         profiles in the best models of KIC\,10526294, using exponential (a) and 
         step-function (b) overshooting prescriptions.
         See Table 4 in MAP15 for the mass and extent of each zone.
         The dashed lines show the Brunt-V\"{a}is\"{a}l\"{a} $N^2_{\rm BV}$ 
         (Eq.\ref{e-brunt}) profiles.
         (c) The observed (grey) versus modelled (black) period spacing 
         from the best seismic model (with $f_{\rm ov}=0.017$).
         }
\label{f-dP-Dmix}
\end{figure}

MAP15 compared the exponentially decaying  
($D_{\rm ov}=D_{\rm conv}\exp[-2(r-r_{\rm core})/f_{\rm ov}H_p)]$) and the
step-function ($D_{\rm ov}=D_{\rm conv}$ for $r_{core}\leq r\leq \alpha_{\rm ov}H_p$) 
overshooting prescriptions in their extensive seismic modelling  of KIC\,10526294.
Note that $\alpha_{\rm ov}\approx10 f_{\rm ov}$. 
The mixing induced by convective overshoot 
is \textit{assumed} diffiusive \citep{zhang-2013-01,viallet-2015-01}
with an exponentially decaying behaviour \citep{herwig-2000-01} (Fig.\ref{f-dP-Dmix}a)
or step-function behaviour (Fig.\ref{f-dP-Dmix}b).
We also \textit{assume} that the thermal stratification in the overshoot layer is 
purely radiative $\nabla=\nabla_{\rm rad}$.
Clearly, $N^2_{\rm BV}$ (red dashed) shows a steeper rise in the step-overshoot model.
Fig.\,1c shows the observed versus modelled $\Delta P$,
where the first five modes agree very well with observations,
and the other thirteen deviate significantly from the fit.
We argue the reason for this is our reasonable ability in confining the width of the 
$\mu$-gradient layer, and at the same time our poor understanding of the thermal and 
chemical stratification in the overshooting layer. 

\section{Mode trapping}\label{s-trapping}
\begin{figure}
\begin{minipage}{0.49\textwidth}
\includegraphics[width=\columnwidth]{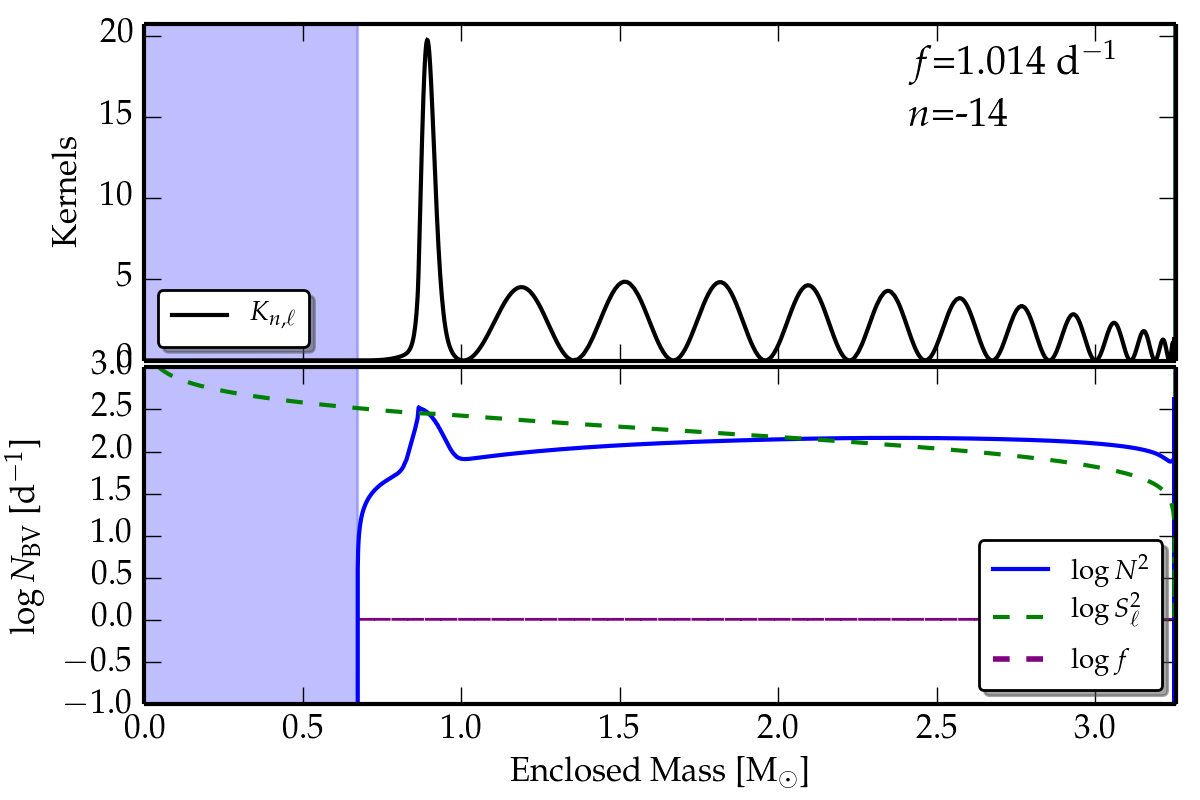}
\end{minipage}
\hspace{0.01\textwidth}
\begin{minipage}{0.49\textwidth}
\includegraphics[width=\columnwidth]{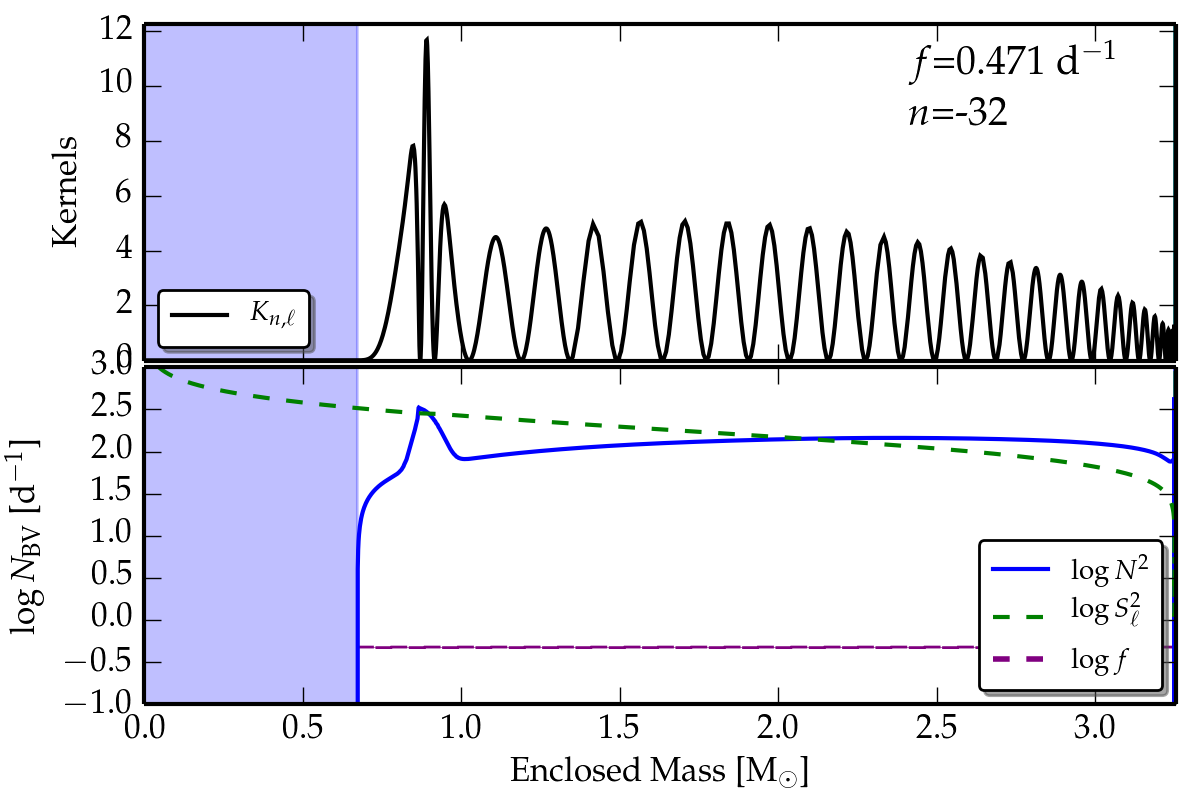}
\end{minipage}
\caption{The rotation kernels for the shortest period mode $n=-14$ (Left)
and for the longest period mode $n=-32$ (Right) chosen from the best model of KIC\,10526294.
The profile of the Brunt-V\"{a}is\"{a}l\"{a} frequency $N_{\rm BV}$ and the Lamb 
frequency are shown to guide the eyes.
The convective core is highlighted in blue.}
\label{f-eigs}
\end{figure}

Fig.\,\ref{f-eigs} presents the behaviour of the rotation kernels $K_{n,\ell}$
(defined in \citet{aerts-2010-book}, Eq.\,3.356) versus enclosed mass.
The bottom panels show the run of the Brunt-V\"{a}is\"{a}l\"{a} frequency $N_{\rm BV}$,
the Lamb frequency $S_{\ell}$ and the mode frequency in logarithmic scale.
The $\mu$-gradeint zone corresponds to the local peak in $N_{\rm BV}$ around 0.85\,M$_\odot$.
The shortest period mode (left) is perfectly trapped in the $\mu$-gradient layer, and
has a sizable amplitude there \citep{dziembowski-1991-01}.
The longest period mode (right) still has a large amplitude in the $\mu$-gradient layer, 
in addition to a significant amplitude in the diffusive oveshooting 
layer below the $\mu$-gradient zone.
Thus, the eigenfunctions of the latter mode are sensitive to the treatment of overshooting
(Figs.\ref{f-dP-Dmix}a and \ref{f-dP-Dmix}b).
The frequencies and period spacings assosicated to highest-order g-modes 
of KIC\,10526294 (Fig.\,\ref{f-dP-Dmix}c) do not perfectly match the observation. 
This manifests that our underlying assumptions about the thermal and compositional
structure of the overshooting layer build in 1D stellar structure codes need a major
revision.

The current assumption in 1D stellar models is that the temperature gradient $\nabla$
changes abruptly from fully adiabatic in the convective core to fully radiative 
just above the convective core, where the convective plumes overshoot into the radiative
regions. 
However, this picture is not supported by 3D simulations of \citet{viallet-2013-01} 
(although performed for the conditions of deep convection in red giant envelopes).
It is possible that the behaviour of the temperature gradient in the overshooting layer
is quasi-adiabatic, up to some pressure scale heights from the core boundary, where it 
efficiently mixes the materials and adds to the size of the fully mixed core. 
Such a prescription is similar to the classical step-function overshoot.
An extension of this quasi-adiabatic layer can be photon-dominated, and an exponentially
decaying diffusive mixing can be applicable there.
Therefore, a more complicated 1D stellar evolution models should be built, hopefully 
with improved treatment of convection \citep{zhang-2013-01,arnett-2015-01} to provide 
better inputs for asteroseismic inferences.
Indeed, the one-to-one frequency matching with observations can discriminate between
the most plausible convective/overshoot prescription.

\textbf{Acknowledgement.} 
EM thanks the support from the Belgian Federal Science Policy Office (BELSPO)
grant n$^\circ$246540, the Marie Curie IIF grant n$^\circ$623303, and the Research 
Council of KU\,Leuven grant GOA/2013/012.

{\footnotesize
\linespread{0.25}
\bibliography{my-bib.bib}
\bibliographystyle{astron}
}

\end{document}